\documentclass[aps,prl,preprint,superscriptaddress]{revtex4}
\usepackage{graphicx}
\usepackage{epstopdf}
\usepackage{amsmath,amssymb,bm,graphics}

\usepackage[usenames]{color}
\usepackage[colorlinks=true, urlcolor=navyblue, linkcolor=navyblue, citecolor=navyblue]{hyperref}
\usepackage{relsize}
\definecolor{navyblue}{rgb}{0,0.08,0.45}

\usepackage{color}

\newcommand{\half}{{\textstyle{\frac{1}{2}}}}

\begin{document}

\preprint{ SLAC-PUB-15377}

\title{Conformal Symmetry,  Confinement, \\
and Light-Front Holographic QCD}

\author{Stanley J. Brodsky}
\affiliation{SLAC National Accelerator Laboratory, Stanford University,
Stanford, California 94309, USA}

\author{Guy F. de T\'eramond}
\affiliation{Universidad de Costa Rica, San Jos\'e, Costa Rica}

\author{Hans G\"unter Dosch}
\affiliation{Institut f\"ur Theoretische Physik, Philosophenweg
16, D-6900 Heidelberg, Germany}


\begin{abstract}  We show that  (a) the conformal properties of Anti-de Sitter (AdS) space, (b)  
the  properties of a field theory in one dimension under the full conformal group
found by  de Alfaro, Fubini and Furlan, and (c) the frame-independent single-variable light-front Schr\"odinger equation for bound states all lead to the same result:  a relativistic nonperturbative model which successfully incorporates salient features of hadronic physics,  including 
confinement, linear Regge trajectories, and results which are conventionally attributed to  spontaneous chiral symmetry breaking.
\end{abstract}

\maketitle

\section{Conformal Invariance and QCD}
One of the most intriguing features of QCD for massless quarks is its underlying conformal invariance, invariance under both scale  (dilatation) and special conformal transformations~\cite{Parisi:1972zy}.
For example, in the case of perturbative QCD, the running coupling  $\alpha_s(Q^2)$ becomes constant in the limit of zero $\beta$-function and zero quark mass, and conformal symmetry becomes manifest.  In fact, the renormalization scale uncertainty in 
pQCD predictions can be eliminated by using the Principle of Maximum Conformality (PMC)~\cite{Brodsky:2011ig}.
Using the PMC/BLM procedure~\cite{Brodsky:1982gc},
all non-conformal contributions in the perturbative expansion series are summed into the running coupling by shifting the renormalization scale in $\alpha_s$ from its initial value, and one obtains unique, scale-fixed, scheme-independent predictions at any finite order.  One can also introduce a generalization of conventional dimensional regularization which illuminates the renormalization scheme and scale ambiguities of pQCD predictions, exposes the general pattern of nonconformal terms, and allows one to systematically determine the argument of the running coupling order by order in pQCD in a form which can be readily automatized~\cite{Mojaza:2012mf}.
The resulting PMC scales and  finite-order PMC predictions are both to high accuracy independent of the choice of initial renormalization scale.  For example, PMC scale-setting leads to a scheme-independent 
pQCD prediction~\cite{Brodsky:2012ik}
for the top-quark forward-backward asymmetry which is within one $\sigma$ of the Tevatron measurements. 
The PMC procedure also provides scale-fixed, scheme-independent commensurate scale relations~\cite{Brodsky:1994eh}, relations between observables which are based on the underlying conformal behavior of QCD such as the Generalized Crewther relation~\cite{Brodsky:1995tb}.
The PMC satisfies  all of the principles of the renormalization group: reflectivity, symmetry, and transitivity, and it thus eliminates an unnecessary source of systematic error in pQCD predictions~\cite{Wu:2013ei}.

Anti-de Sitter space in five dimensions (AdS$_5$) provides a geometric representation of the conformal group. One can modify AdS space by using a dilaton factor in the AdS metric 
$e^{\varphi(z)} $ to introduce the QCD 
confinement scale. Although such a mass scale explicitly breaks the dilatation invariance of the equations of motion, the action can still be conformally invariant, as was first shown   by  V.~de Alfaro, S.~Fubini and G.~Furlan (dAFF)~\cite{deAlfaro:1976je} in the context of one-dimensional quantum 
field theory;  i.e, the change in the mass scale of the potential can be compensated by the time scale in the action.  However, this manifestation of conformal symmetry can only occur if  the dilaton profile  $\varphi(z) \propto z^s$ is constrained to have the specific power $s = 2$,  a remarkable result which follows from the dAFF construction of conformally invariant quantum mechanics~\cite{Brodsky:2013kpr}.     The quadratic form $\varphi(z) =\kappa^2 z^2$ leads to linear Regge trajectories~\cite{Karch:2006pv} in the hadron mass squared.   

\section{Light-Front Holography}

A remarkable holographic feature of dynamics in AdS space in five dimensions is that it is dual to Hamiltonian theory in physical space-time, 
quantized at fixed light-front (LF) time~\cite{deTeramond:2008ht}.  For example, the equation of motion for mesons on the light-front has exactly the same single-variable form as the AdS/QCD equation of motion; one can then interpret the AdS fifth dimension variable $z$ in terms of the physical variable $\zeta$, representing the invariant separation of the $q$ and $\bar q$ at fixed light-front time.  As discussed in the next sections, this light-front holographic principle provides a precise relation between the bound-state amplitudes in AdS space and the boost-invariant LF wavefunctions describing the internal structure of hadrons in physical space-time.  The resulting valence Fock-state wavefunction eigensolutions of the LF QCD Hamiltonian satisfy a single-variable relativistic equation of motion analogous to the nonrelativistic radial Schr\"odinger equation. 
The quadratic dependence in the quark-antiquark potential 
$U(\zeta^2,J) =  \kappa^4 \zeta^2 +2 \kappa^2(J-1)$ 
in the Light-Front Schr\"odinger equation (LFSE) is determined uniquely from conformal invariance, whereas the constant term $ 2 \kappa^2(J-1)$  is fixed by the duality between AdS and LF quantization, a correspondence which follows specifically from the separation of kinematics and dynamics on the light-front~\cite{deTeramond:2013it}.  
The LF potential thus has a specific power dependence--in effect, it is a light-front harmonic oscillator potential.
It  is confining and reproduces the observed linear Regge behavior of the light-quark hadron spectrum in both the orbital angular momentum $L$ and the radial node number $n$. The  pion is predicted to be massless in the chiral limit ~\cite{deTeramond:2009xk}  - the positive contributions to $m^2_\pi$ from the LF potential and kinetic energy is cancelled by the constant term  in $U(\zeta^2,J) $ for $J=0.$  The  derived running QCD coupling displays an infrared fixed point~\cite{Brodsky:2010ur}.

The construction of  dAFF  retains conformal invariance of the action despite the presence of a fundamental mass scale.  
The AdS approach, however, goes beyond the purely group-theoretical considerations of dAFF, since 
features such as the masslessness of the pion and the separate dependence on $J$ and $L$ are  consequences of the potential derived from the holographic LF duality with AdS for general  $J$ and $L$~\cite{Brodsky:2013kpr, deTeramond:2013it}.

The quantization of QCD at fixed light-front time~\cite{Dirac:1949cp} (Dirac's Front Form)  provides a first-principles Hamiltonian method for solving nonperturbative QCD. It is rigorous, has no fermion-doubling, is formulated in Minkowski space, and it is frame-independent.  Given the boost-invariant light-front wavefunctions $\psi_{n/H}$ (LFWFs), one can
compute a large range of hadron
observables, starting with structure functions, generalized parton distributions, and form factors.
It is also possible to compute jet hadronization at the amplitude level from first principles from the LFWFs~\cite{Brodsky:2008tk}. A similar method has been used to predict the production of antihydrogen from the off-shell coalescence of relativistic antiprotons and positrons~\cite{Munger:1993kq}.
The LFWFs of hadrons thus provide a direct connection between observables and the QCD Lagrangian. 
Solving nonperturbative QCD is thus equivalent to solving the light-front Heisenberg matrix eigenvalue problem.   Angular momentum $J^z$ is conserved at every vertex.
The LF vacuum is defined as the state of lowest invariant mass and is trivial up to zero modes.  There are thus no quark or gluon vacuum condensates in the LF vacuum-- the corresponding physics is contained within the 
LFWFs themselves~\cite{Brodsky:2009zd,  Brodsky:2010xf}, thus eliminating a major contribution to the cosmological constant.

The simplicity of the front form contrasts with the usual instant-form formalism. Current matrix elements defined at ordinary time $t$ must include the coupling of photons and vector bosons fields  to connected vacuum-induced currents; otherwise, the result is not Lorentz-invariant.  Thus the knowledge of the hadronic eigensolutions of the instant-form Hamiltonian are insufficient for determining form factors or other observables.   In addition, the boost of an instant form wavefunction from $p$ to $p+q$ changes particle number and is an extraordinarily complicated dynamical problem. 

It is remarkable fact that AdS/QCD,  which was originally motivated by
the AdS/CFT correspondence between gravity on a higher-dimensional 
space and conformal field theories 
in physical space-time~\cite{Maldacena:1997re},  has a direct holographic mapping to light-front Hamiltonian theory~\cite{deTeramond:2008ht}. 
The AdS mass  parameter $\mu R$ maps to the LF orbital angular momentum.  The formulae for electromagnetic~\cite{Polchinski:2002jw} and gravitational~\cite{Abidin:2008ku} form factors in AdS space map to the exact Drell-Yan-West formulae in light-front QCD~\cite{Brodsky:2006uqa, Brodsky:2007hb, Brodsky:2008pf}.  Thus the light-front holographic approach provides an analytic frame-independent first approximation to the color-confining dynamics,  spectroscopy, and excitation spectra of the relativistic light-quark bound states of QCD.   It is systematically improvable in full QCD using  the basis light-front quantization (BLFQ) method~\cite{Vary:2009gt} and other methods.

\section{The Light-Front Schr\"odinger Equation: A Semiclassical Approximation to QCD \label{LFQCD}}

In the limit of zero quark masses the longitudinal modes decouple  from the  invariant  LF Hamiltonian  equation  $H_{LF} \vert \phi \rangle  =  M^2 \vert \phi \rangle$
with  $H_{LF} = P_\mu P^\mu  =  P^- P^+ -  \vec{P}_\perp^2$. The generators $P = (P^-, P^+,  \vec{P}_\perp)$, $P^\pm = P^0 \pm P^3$,  are constructed canonically from the QCD Lagrangian by quantizing the system on the light-front at fixed LF time $x^+$, $x^\pm = x^0 \pm x^3$~\cite{Brodsky:1997de}. The LF Hamiltonian $P^-$ generates the LF time evolution with respect to $x^+$,
whereas the LF longitudinal $P^+$ and transverse momentum $\vec P_\perp$ are kinematical generators.

It is advantageous to reduce the full multiparticle eigenvalue problem of the LF Hamiltonian to an effective light-front Schr\"odinger equation  which acts on the valence sector LF wavefunction and determines each eigensolution separately~\cite{Pauli:1998tf}.   In contrast,  diagonalizing the LF Hamiltonian yields all eigensolutions simultaneously, a complex task.
The central problem for deriving the LFSE becomes the derivation of the effective interaction $U$ which acts only on the valence sector of the theory and has, by definition, the same eigenvalue spectrum as the initial Hamiltonian problem.  In order to carry out this program one must systematically express the higher Fock components as functionals of the lower ones. This  method has the advantage that the Fock space is not truncated, and the symmetries of the Lagrangian are preserved~\cite{Pauli:1998tf}.

The light-front Hamiltonian for QCD can be derived directly from the QCD Lagrangian~\cite{Brodsky:1997de}.
The result is relativistic and frame-independent.    The $q \bar q$ LF Fock state wavefunction for a meson  can be written as
$\psi(x,\zeta, \varphi) = e^{i L \theta} X(x) \frac{\phi(\zeta)}{\sqrt{2 \pi \zeta}},$
thus factoring the angular dependence $\theta$ and the longitudinal, $X(x)$, and transverse mode $\phi(\zeta)$.
In the limit of zero quark masses the longitudinal mode decouples and
the LF eigenvalue equation $P_\mu P^\mu \vert \phi \rangle  =  M^2 \vert \phi \rangle$
takes the form of a light-front  wave equation for $\phi$~\cite{deTeramond:2008ht}
\begin{equation} \label{LFWE}
\left[-\frac{d^2}{d\zeta^2}
- \frac{1 - 4L^2}{4\zeta^2} + U\left(\zeta^2, J, M^2\right) \right]
\phi_{J,L,n}(\zeta^2) = M^2 \phi_{J,L,n}(\zeta^2),
\end{equation}
a relativistic {\it single-variable}  LF  Schr\"odinger equation. 
This equation describes the spectrum of mesons as a function of $n$, the number of nodes in $\zeta$, the total angular momentum  $J$, which represent the maximum value of $\vert J^z \vert$, $J = \max \vert J^z \vert$,
and the internal orbital angular momentum of the constituents $L= \max \vert L^z\vert$.
The variable $z$ of AdS space is identified with the LF   boost-invariant transverse-impact variable $\zeta$~\cite{Brodsky:2006uqa}, 
thus giving the holographic variable a precise definition in LF QCD~\cite{deTeramond:2008ht, Brodsky:2006uqa}.
For a two-parton bound state $\zeta^2 = x(1-x) b^{\,2}_\perp$,
where $x$ is the longitudinal momentum fraction and $ b_\perp$ is  the transverse-impact distance between the quark and antiquark. 
In the exact QCD theory $U$ is related to the two-particle irreducible $q \bar q$ Green's function.

The potential in the LFSE is determined from the two-particle irreducible (2PI) $ q \bar q \to q \bar q $ Greens' function.  In particular, the higher Fock states in intermediate states
leads to an effective interaction $U(\zeta^2, J ,M^2)$  for the valence $\vert q \bar q \rangle$ Fock state~\cite{Pauli:1998tf}.
A related approach for determining the valence light-front wavefunction and studying the effects of higher Fock states without truncation has been given in Ref.~\cite{Chabysheva:2011ed}.

Unlike ordinary instant-time quantization, the light-front Hamiltonian equations of motion are frame independent; remarkably, they  have a structure which matches exactly the eigenmode equations in AdS space. This makes a direct connection of QCD with AdS methods possible.  In fact, one can
derive the light-front holographic duality of AdS  by starting from the light-front Hamiltonian equations of motion for a relativistic bound-state system
in physical space-time~\cite{deTeramond:2008ht}.

\section{Effective Confinement from the Gauge/Gravity Correspondence}

Recently we have derived wave equations for hadrons with arbitrary spin starting from an effective action in  AdS space~\cite{deTeramond:2013it}.    An essential element is the mapping of the higher-dimensional equations  to the LF Hamiltonian equation  found in Ref.~\cite {deTeramond:2008ht}.  This procedure allows a clear distinction between the kinematical and dynamical aspects of the LF holographic approach to hadron physics.  Accordingly, the non-trivial geometry of pure AdS space encodes the kinematics,  and the additional deformations of AdS encode the dynamics, including confinement~\cite{deTeramond:2013it}.

A spin-$J$ field in AdS$_{d+1}$ is represented by a rank $J$ tensor field $\Phi_{M_1 \cdots M_J}$, which is totally symmetric in all its indices.  In presence of a dilaton background field $\varphi(z)$ the effective action is~\cite{deTeramond:2013it} 
\begin{multline}
\label{Seff}
S_{\it eff} = \int d^{d} x \,dz \,\sqrt{\vert g \vert}  \; e^{\varphi(z)} \,g^{N_1 N_1'} \cdots  g^{N_J N_J'}   \Big(  g^{M M'} D_M \Phi^*_{N_1 \dots N_J}\, D_{M'} \Phi_{N_1 ' \dots N_J'}  \\
 - \mu_{\it eff}^2(z)  \, \Phi^*_{N_1 \dots N_J} \, \Phi_{N_1 ' \dots N_J'} \Big),
 \end{multline}
where the indices $M, N = 0, \cdots , d$, $\sqrt{g} = (R/z)^{d+1}$ and $D_M$ is the covariant derivative which includes parallel transport. The coordinates of AdS are the Minkowski coordinates $x^\mu$ and the holographic variable $z$, $x^M = \left(x^\mu, z\right)$. 
The effective mass  $\mu_{\it eff}(z)$, which encodes kinematical aspects of the problem, is an {\it a priori} unknown function,  but the additional symmetry breaking due to its $z$-dependence allows a clear separation of kinematical and dynamical effects~\cite{deTeramond:2013it}.
The dilaton background field $\varphi(z)$ in  (\ref{Seff})   introduces an energy scale in the five-dimensional AdS action, thus breaking conformal invariance. It  vanishes in the conformal ultraviolet limit $z \to 0$.

 A physical hadron has plane-wave solutions and polarization indices along the 3 + 1 physical coordinates
 $\Phi_P(x,z)_{\nu_1 \cdots \nu_J} = e^{ i P \cdot x} \Phi_J(z) \epsilon_{\nu_1 \cdots \nu_J}({P})$,
 with four-momentum $P_\mu$ and  invariant hadronic mass  $P_\mu P^\mu \! = M^2$. All other components vanish identically. 
 The wave equations for hadronic modes follow from the Euler-Lagrange equation for tensors orthogonal to the holographic coordinate $z$,  $\Phi_{z N_2 \cdots N_J}  = 0$. Terms in the action which are linear in tensor fields, with one or more indices along the holographic direction, $\Phi_{z N_2 \cdots N_J}$, give us 
 the kinematical constraints required to eliminate the lower-spin states~\cite{deTeramond:2013it}.  Upon variation with respect to $ \hat \Phi^*_{\nu_1 \dots \nu_J}$,
 we find the equation of motion~\cite{deTeramond:2013it}  
\begin{equation}  \label{PhiJM}
 \left[ 
   -  \frac{ z^{d-1- 2J}}{e^{\varphi(z)}}   \partial_z \left(\frac{e^{\varphi(z)}}{z^{d-1-2J}} \partial_z   \right) 
  +  \frac{(m\,R )^2}{z^2}  \right]  \Phi_J = M^2 \Phi_J,
  \end{equation}
  with  $(m \, R)^2 =(\mu_{\it eff}(z) R)^2  - J z \, \varphi'(z) + J(d - J +1)$,
  which is  the result found in Refs.~\cite{deTeramond:2008ht, deTeramond:2012rt} by rescaling the wave equation for a scalar field. Similar results were found
  in Ref.~\cite{Gutsche:2011vb}.
 Upon variation with respect to
$ \hat \Phi^*_{N_1 \cdots z  \cdots N_J}$  we find the kinematical constraints which  eliminate lower spin states from the symmetric field tensor~\cite{deTeramond:2013it}  
\begin{equation} \label{sub-spin}
 \eta^{\mu \nu } P_\mu \,\epsilon_{\nu \nu_2 \cdots \nu_J}({P})=0, \quad
\eta^{\mu \nu } \,\epsilon_{\mu \nu \nu_3  \cdots \nu_J}({P})=0.
 \end{equation}

Upon the substitution of the holographic variable $z$ by the LF invariant variable $\zeta$ and replacing
  $\Phi_J(z)   = \left(R/z\right)^{J- (d-1)/2} e^{-\varphi(z)/2} \, \phi_J(z)$ 
in (\ref{PhiJM}), we find for $d=4$ the LFSE (\ref{LFWE}) with effective potential~\cite{deTeramond:2010ge}
\begin{equation} \label{U}
U(\zeta^2, J) = \frac{1}{2}\varphi''(\zeta^2) +\frac{1}{4} \varphi'(\zeta^2)^2  + \frac{2J - 3}{2 \zeta} \varphi'(\zeta^2) ,
\end{equation}
provided that the AdS mass $m$ in (\ref{PhiJM}) is related to the internal orbital angular momentum $L = max \vert L^z \vert$ and the total angular momentum $J^z = L^z + S^z$ according to $(m \, R)^2 = - (2-J)^2 + L^2$.  The critical value  $L=0$  corresponds to the lowest possible stable solution, the ground state of the LF Hamiltonian.
For $J = 0$ the five dimensional mass $m$
 is related to the orbital  momentum of the hadronic bound state by
 $(m \, R)^2 = - 4 + L^2$ and thus  $(m\, R)^2 \ge - 4$. The quantum mechanical stability condition $L^2 \ge 0$ is thus equivalent to the Breitenlohner-Freedman stability bound in AdS~\cite{Breitenlohner:1982jf}.

\begin{figure}[h]
\centering
\includegraphics[width=6.15cm]{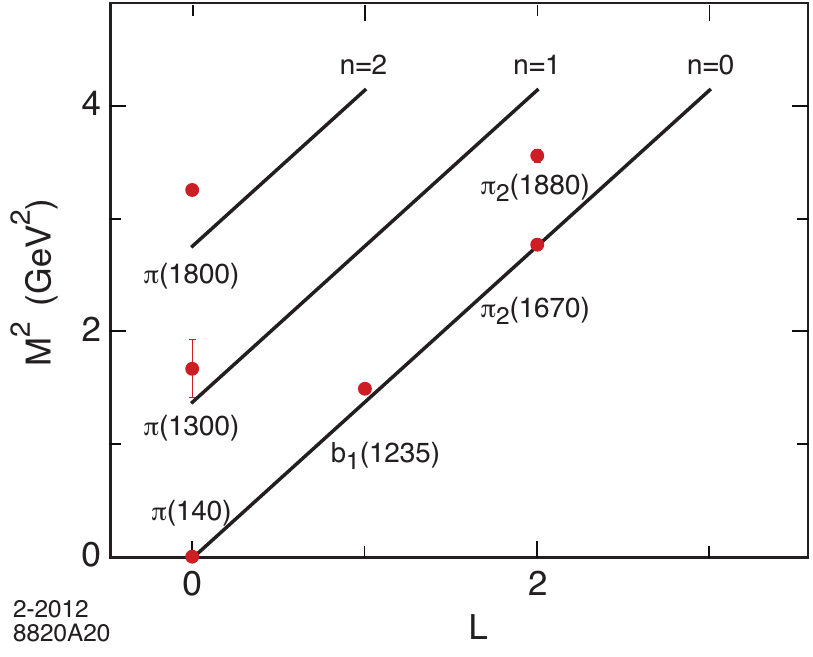} \hspace{0pt}
\includegraphics[width=6.15cm]{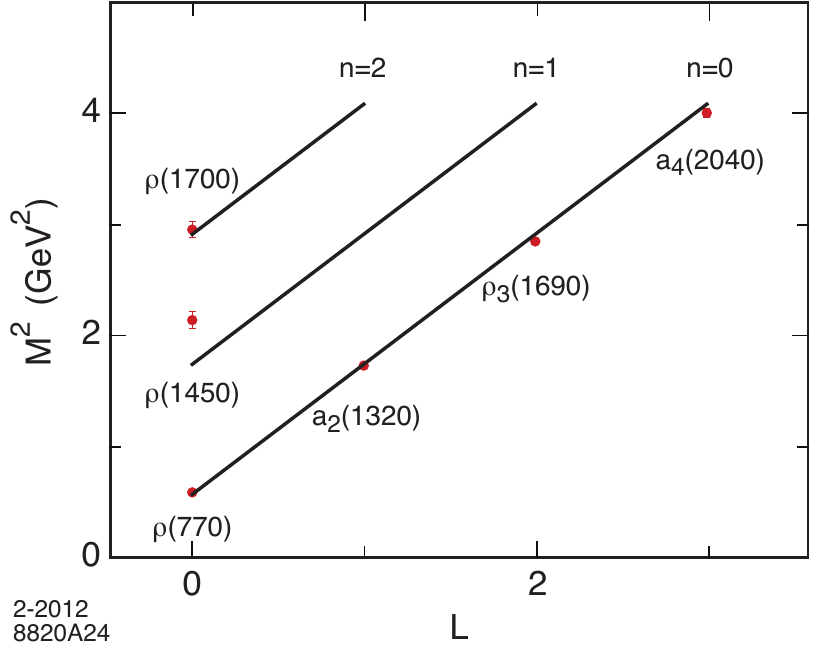}
 \caption{$I = 1$ parent and daughter Regge trajectories for the $\pi$-meson family (left) with
$\kappa= 0.59$ GeV; and  the   $\rho$-meson
 family (right) with $\kappa= 0.54$ GeV.}
\label{pionspec}
\end{figure}

 A particularly interesting example is a dilaton profile $\exp{\left(\pm \kappa^2 z^2\right)}$ of either sign, since it 
leads to linear Regge trajectories~\cite{Karch:2006pv} and avoids the ambiguities in the choice of boundary conditions at the infrared wall.  
For the  confining solution $\varphi = \exp{\left(\kappa^2 z^2\right)}$ the effective potential is
$U(\zeta^2,J) =   \kappa^4 \zeta^2 + 2 \kappa^2(J - 1)$ and  Eq.  (\ref{LFWE}) has eigenvalues
$M_{n, J, L}^2 = 4 \kappa^2 \left(n + \frac{J+L}{2} \right)$,
with a string Regge form $M^2 \sim n + L$.  
A discussion of the light meson and baryon spectrum,  as well as  the elastic and transition form factors of the light hadrons using LF holographic methods, is given in 
Ref.~\cite{deTeramond:2012rt}.  As an example the spectral predictions  for the $J = L + S$ light pseudoscalar and vector meson  states are  compared with experimental data in Fig. \ref{pionspec} for the positive sign dilaton model.

The effective interaction $U(\zeta^2,J)$
is instantaneous in LF time and acts on the lowest state of the LF Hamiltonian.  This equation describes the spectrum of mesons as a function of $n$, the number of nodes in $\zeta^2$,
 the internal orbital angular momentum $L = L^z$, and the total angular momentum $J=J^z$,
with $J^z = L^z + S^z$  the sum of the  orbital angular momentum of the constituents and their internal spins.
[The  $SO(2)$ Casimir  $L^2$  corresponds to  the group of rotations in the transverse LF plane.]
It is the relativistic frame-independent front-form analog of the non-relativistic radial Schr\"odinger equation for muonium  and other hydrogenic atoms in presence of an instantaneous Coulomb potential.

The AdS/QCD harmonic oscillator potential could in fact emerge from the exact QCD formulation when one includes contributions from the LFSE potential $U$ which are due to the exchange of two connected gluons; {\it i.e.}, ``H'' diagrams~\cite{Appelquist:1977tw}.
We notice that $U$ becomes complex for an excited state since a denominator can vanish; this gives a complex eigenvalue and the decay width.

The correspondence between the LF and AdS equations  thus determines the effective confining interaction $U$ in terms of the infrared behavior of AdS space and gives the holographic variable $z$ a kinematical interpretation. The identification of the orbital angular momentum 
is also a key element of our description of the internal structure of hadrons using holographic principles.

\section{Uniqueness of the Confining Potential}

If one starts with a dilaton profile  $e^{\varphi(z)}$ with $\varphi \propto z^s,$  the existence of a massless pion in the limit of massless quarks determines
uniquely the value  $s = 2$.
To show this, one can use the stationarity of bound-state energies with respect to variation of parameters~\cite{Brodsky:2013kpr}.
The quadratic dilaton profile also follows  from
the  algebraic construction of Hamiltonian operators  by (dAFF)~\cite{deAlfaro:1976je}.
The action
$
{ \cal{S}}= \half \int dt (\dot Q^2 - g/Q^2),
$
 is invariant under conformal transformations in the 
variable $t$, and  there are 
in addition to the Hamiltonian  $H_t$ two more invariants of motion for this field theory, namely the 
dilation operator $D$ and $K$, corresponding to the special conformal transformations in $t$.  
Specifically, if one introduces the  the new variable $\tau$ defined through 
$d\tau= d t/(u+v\,t + w\,t^2)$ and the  rescaled fields $q(\tau) = Q(t)/(u + v\, t + w \,t^2)^{1/2}$,
it then follows that the the operator
$G= u\,H_t + v\,D + w\,K$  generates
evolution in  $\tau$~\cite{deAlfaro:1976je}.
The Hamiltonian corresponding to the operator $G$ which introduces the mass scale
is a linear combination of the old Hamiltonian $H_t$, $D$, the generator of dilations, and
$K$, the generator of special conformal transformations.
It contains the confining potential 
$(4 u w - v^2) \zeta^2/8$, that is the confining term in (\ref{U}) for a quadratic dilaton profile and thus $\kappa^4 =  (4 u w - v^2)/8$.

The construction of new Hamiltonians from the generators of the conformal group has been used by 
dAFF to construct algebraically the spectra and the eigenfunctions of these operators. The  conformal group in one dimension is locally isomorphic to the group of pseudo-rotations $O(2,1)$. The compact operator $R= \frac{1}{2} (K + H_t)$ generates rotations in the Euclidean 1-2 plane,  whereas   the non-compact operators
 $L_1=\frac{1}{2} (K-H_t)$  and $L_2=D$ generate  pseudo-rotations (boosts) in the non-Euclidean 2-3 and 1-3 plane respectively.  As in the familiar case of angular momentum, one can introduce raising and lowering operators $L_\pm = L_1 \pm L_2$ and construct the spectrum and the eigenfunctions analogously to the 
 angular momentum operators.  Interestingly, this is just the method employed in Ref.~\cite{Brodsky:2008pg} 
 to obtain an integrable  LF confining  Hamiltonian by following Infeld's observation that integrability follows immediately if the equation of motion can be factorized as a product of linear operators~\cite{Infeld:1941}. The method can be extended to describe baryons in AdS while preserving the algebraic structure~\cite{Brodsky:2008pg, MartinezyMoreno:1992zz}.  
 Our approach has elements in common with those of Ref.~\cite{Hoyer:2012vm}, where the scale of the confinement potential arises from a boundary condition when solving Gauss' equation.

\section{Summary}

The triple complementary connection of  (a)  AdS space,  (b) its LF holographic dual, 
and (c) the relation to the algebra of the conformal group in one dimension, is characterized by a quadratic confinement LF potential, and 
thus a dilaton profile with the power $z^s$, with the unique power $s = 2$. In fact,  for $s=2$ the mass of the $J=L=n=0$  pion is 
automatically zero in the chiral limit, and the separate dependence on $J$ and $L$ leads to a  mass ratio of the $\rho$ and the $a_1$ mesons which coincides with the result of the Weinberg sum rules~\cite{Weinberg:1967kj}. One predicts linear  Regge trajectories  with the same slope in the relative orbital angular momentum $L$ and the principal quantum humber $n$.  The AdS approach, however,  goes beyond the purely group theoretical considerations of dAFF, since 
features such as the masslessness of the pion and the separate dependence on $J$ and $L$ are a consequence of the potential (\ref{U}) derived from the duality with AdS for general $J$ and $L$.
The constant term in the potential, which is not determined by the group theoretical arguments, is  fixed by the holographic duality to LF  quantized QCD~\cite{deTeramond:2013it}.
The resulting Lagrangian, constrained by the conformal invariance of the 
action, has the same form as the AdS Lagrangian with a quadratic dilaton profile.

In their discussion of the evolution operator $H_\tau$ 
as a model for confinement, 
dAFF mention a critical point, namely that ``the time evolution is quite different from a stationary one''.  By this statement they refer to the fact that the variable $\tau$ is related to the variable $t$ for the case  $u w >0, \,v=0$   by 
$
\tau =\frac{1}{\sqrt{u\,w}} \arctan\left(\sqrt{\frac{w}{u}} t\right),
$
{\it i.e.}, $\tau$ has only a limited range. The finite range of invariant LF time $\tau=x^+/P^+$ can be interpreted as a feature of the internal frame-independent LF  time difference between the confined constituents in a bound state. For example, in the collision of two mesons, it would allow us to compute the LF time difference between the two possible quark-quark collisions.

The treatment of the chiral limit in the LF holographic approach to strongly coupled QCD is substantially different from the standard approach  based on chiral perturbation theory.
In the conventional approach \cite{Goldstone:1962es}, spontaneous symmetry breaking  by a  non-vanishing chiral quark condensate $\langle  \bar \psi \psi  \rangle$ plays the crucial role.  In QCD sum  rules \cite{Shifman:1978bx}  $\langle  \bar \psi \psi \rangle$ brings in non-perturbative elements into the perturbatively calculated spectral sum rules. It should  be noted, however, that the definition of the condensate, even in lattice QCD necessitates a renormalization procedure for the operator product, and it is not a directly observable quantity.  In Bethe-Salpeter~\cite{Maris:1997hd} and light-front analyses~\cite{Brodsky:2012ku}, the Gell Mann-Oakes-Renner relation~\cite{GellMann:1968rz}
for $m^2_\pi/m_q$ involves the decay matrix element $\langle 0 |\bar \psi\gamma_5 \psi |\pi \rangle$ instead of $\langle 0| \bar \psi \psi | 0\rangle$.

In the color-confining AdS/QCD light-front model discussed here, the vanishing of the pion mass in the chiral limit, a phenomenon usually ascribed to spontaneous symmetry breaking of the chiral symmetry,  is  obtained specifically from the precise cancellation of the LF kinetic energy and LF potential energy terms for the quadratic confinement potential. This mechanism provides a  viable alternative to the conventional description of nonperturbative QCD based on vacuum condensates, and it 
eliminates a major conflict of hadron physics with the empirical value for the cosmological constant~\cite{Brodsky:2009zd,  Brodsky:2010xf}.

\section{acknowledgments}
Invited talk, presented by SJB at the Third Workshop on the QCD Structure of the Nucleon (QCD-N'12), Bilbao, Spain, October 22-26, 2012.
This work  was supported by the Department of Energy contract DE--AC02--76SF00515.

$ $

\end{document}